\documentclass[letterpaper]{article} 
\usepackage{aaai24}  
\usepackage{times}  
\usepackage{helvet}  
\usepackage{courier}  
\usepackage[hyphens]{url}  
\usepackage{graphicx} 
\usepackage{natbib}  
\usepackage{caption} 
\usepackage{algorithm}
\usepackage{algorithmic}

\usepackage{newfloat}
\usepackage{listings}
\DeclareCaptionStyle{ruled}{labelfont=normalfont,labelsep=colon,strut=off} 
\floatstyle{ruled}
\newfloat{listing}{tb}{lst}{}
\floatname{listing}{Listing}
\usepackage{xcolor}
\newcommand{\answerYes}[1]{\textcolor{blue}{#1}} 
 
\newcommand{\answerNA}[1]{\textcolor{gray}{#1}}

\usepackage{enumerate}
\usepackage{enumitem}
\usepackage{caption}
\usepackage{subcaption}
\usepackage{booktabs}
\usepackage{xcolor}

\newenvironment{packed_item}{
\begin{itemize}[leftmargin=.2in]
  \setlength{\itemsep}{1pt}
  \setlength{\parskip}{0pt}
  \setlength{\parsep}{0pt}
}{\end{itemize}}

\title{How Does Empowering Users with Greater System Control Affect News Filter Bubbles?}

\author {
    Ping Liu\textsuperscript{\rm 1},
    Karthik Shivaram\textsuperscript{\rm 2},
    Aron Culotta\textsuperscript{\rm 2},
    Matthew Shapiro\textsuperscript{\rm 3},
    Mustafa Bilgic\textsuperscript{\rm 4}
}
\affiliations {
    \textsuperscript{\rm 1} LinkedIn Corporation, Sunnyvale, CA USA \\
    \textsuperscript{\rm 2} Department of Computer Science, Tulane University, New Orleans, LA USA\\
    \textsuperscript{\rm 3} Department of Social Sciences, Illinois Institute of Technology, Chicago, IL USA\\
    \textsuperscript{\rm 4} Department of Computer Science, Illinois Institute of Technology, Chicago, IL USA\\
    piliu@linkedin.com, kshvaram@tulane.edu, aculotta@tulane.edu, shapiro@iit.edu, mbilgic@iit.edu
}

\begin{document}

\maketitle

\begin{abstract}
    While recommendation systems enable users to find articles of interest, they can also create ``filter bubbles'' by presenting content that reinforces users' pre-existing beliefs. Users are often unaware that the system placed them in a filter bubble and, even when aware, they often lack direct control over it. To address these issues, we first design a political news recommendation system augmented with an enhanced interface that exposes the political and topical interests the system inferred from user behavior. This allows the user to adjust the recommendation system to receive more articles on a particular topic or presenting a particular political stance. We then conduct a user study to compare our system to a traditional interface and found that the transparent approach helped users realize that they were in a filter bubble. Additionally, the enhanced system led to less extreme news for most users but also allowed others to move the system to more extremes. Similarly, while many users moved the system from extreme liberal/conservative to the center, this came at the expense of reducing political diversity of the articles shown. These findings suggest that, while the proposed system increased awareness of the filter bubbles, it had heterogeneous effects on news consumption depending on user preferences.
\end{abstract}

\section{Introduction}
\label{sec.introduction}

Personalized recommendation systems help users find items of interest and foster new connections~\cite{guo2017deepfm,tang2013social},
but emerging research suggests that there are unintended side-effects. This is particularly the case for systems recommending political content, resulting in ``filter bubbles'' in which users are being pushed toward homogeneous rather than diverse political content~\cite{pariser2011filter,bakshy2015exposure,robertson2021engagement,liu2021interaction}.

Recent research has attempted to quantify filter bubbles and mitigate them algorithmically \cite{masrour2020bursting,liu2021interaction,shivaram2022reducing}; yet, there have been few attempts
to investigate the effects of giving users greater control over the recommendation algorithm. We design and study a recommendation system with two enhancements: (1) \textit{transparency:} the system exposes the current state of the recommendation system, revealing the political and topical interests inferred from user behavior; (2) \textit{interaction:} the interface allows the user to adjust the recommendation system to receive more articles of a particular topic or political stance. Figure~\ref{fig.transp.inter} shows a sample of the enhanced interface, where sliders can be adjusted by users in order to modify the news content that they receive.

The goal of this paper is to understand the impact of such an interface on system behavior and filter bubbles, as compared with the more limited types of interaction allowed by traditional recommendation systems.
We are particularly interested in how the system affects both the diversity of recommended political news articles as well as user engagement with the system. To this end, we conducted a user study of 102 users (recruited from 850 users who completed a demographic survey and political qualification questionnaire), half of whom used the enhanced interface (the treatment group), and half of whom used a more conventional recommendation system where the only available actions were up-vote or down-vote an article (the control group).  We compute measures over the attributes of the top recommended articles at the beginning and end of each session to study how this interface influences the types of articles shown to the user, how those articles change over time, and whether the system is accurate in its recommendations. By analyzing the results of over 3,000 user interactions with these systems, we make the following observations:

\begin{figure}[t]
  \centering
    \includegraphics[width=0.98\linewidth]{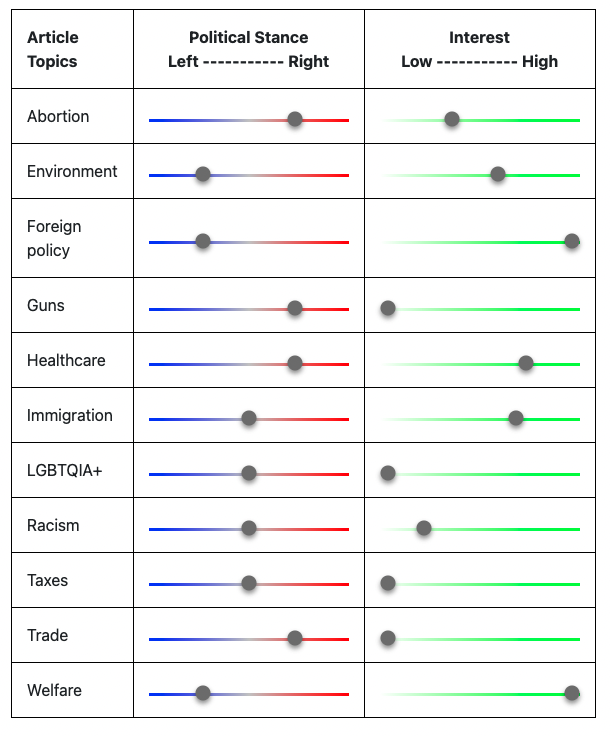}
    \caption{User interface to provide transparency and control over a political news recommendation system.}
    \label{fig.transp.inter}
\end{figure}
\

\vspace{-.1cm}

\begin{packed_item}
\item \textbf{Extremeness:} Among users who were initially exposed to extreme, partisan articles, those in the treatment group were more likely to steer the system to less extreme articles. On the other hand, among users who were initially exposed to less extreme articles, those in the treatment group were somewhat more likely to steer the system to extreme articles. These results suggest a potential extremeness ``sweet spot'' that users seek.
\end{packed_item}

\begin{packed_item}
 \item \textbf{Diversity:} For both treatment and control groups, users  steered the system toward less politically diverse news. The largest difference between the groups was for users who were initially exposed to moderately diverse articles --- such users in the control group steered the system to less diverse articles than users in the treatment group.
\item \textbf{Up-vote Ratio:}
For both treatment and control groups, the ratio of articles that are of up-voted by the users increased over time, particularly for users with low initial up-vote ratios. For users with moderate initial up-vote ratios, those in the treatment group were able to adjust the system to achieve greater system accuracy than those in the control group.
\item \textbf{User awareness:} Through a post-study questionnaire to identify users' motivations and preferences toward our novel recommender system interface, we observed that the transparency of the enhanced interface raised user awareness regarding both the lack of diversity in their recommended articles as well as the inner-workings of news recommendation systems.
\end{packed_item}

The rest of the paper is organized as follows. We discuss the relevant literature and our research questions in \S\ref{sec.related_work}, we detail our approach in \S\ref{sec.details}, followed by a presentation of the findings of the user studies in
\S\ref{sec.findings}. We then discuss limitations of the present work as well as the potential for future research in \S\ref{sec.discussion}, followed by a concluding section.

\section{Related Work and Research Questions}
\label{sec.related_work}

Building on pioneering research on news recommender systems by \citet{chesnais-95}, \citet{kamba1995}, and \citet{claypool1999}, \citet{mitova2023} provide a systematic survey and investigation of news recommender systems in terms of how they affect journalists/media outlets (delivery perspective) and news readers (acquisition perspective). A central challenge posed by recommender systems is the lack of diversity in the recommended items \cite{kunaver2017diversity}. Among the reasons for this diversity problem are: bias introduced by content (e.g., the model latching on to specific keywords), feedback loops when recommendations by the model end up in its training data, and popularity bias when popular rather than niche items are recommended. These are long-standing concerns and have been studied as early as by \citet{smyth-2001} and \citet{ziegler2005}.

The lack of recommendation diversity can create a negative experience for users, as they may be exposed to similar content repeatedly while missing niche content about books, movies, and consumer goods. In the news recommendation domain, it can also lead to echo chambers and filter bubbles \cite{pariser2011filter}, where users are overrecommended news items --- particularly political news --- with which they are ideologically or otherwise aligned. When exposed to political content consistent with one's views, people typically prefer more of the same~\cite{rodriguez2017partisan}. The feedback loops created by these kinds of recommendation models exacerbates echo chambers~\cite{pariser2011filter,bakshy2015exposure,liu2021interaction}, with varying effects across parties~\cite{tewksbury2015polarization}.

User feedback and engagement is critical for understanding the mechanisms underlying filter bubbles. \citet{munson-chi10} conducted a user study where people were assigned either ideologically consistent or inconsistent recommendations and asked to rate them. When the recommendations were unaligned, user satisfaction was low; however, when the list contained a large percentage of agreeable items, responses were much varied: some users were more satisfied, while others were not, suggesting that some people can be ``challenge-averse'' while others are ``diversity-seeking.'' In a similar vein, \citet{liu2021interaction} curated a political news dataset covering numerous topics and conducted simulations using content-based and collaborative-filtering recommender systems. Users who were initially presented extreme news were subsequently presented even more extreme news, users shown more extreme news had higher up-vote ratios, and the recommender system had the least recommendation accuracy for users with diverse views among the various news topics, often resulting in recommendations that were ideologically uniform across topics. This is relevant given that exposure to diverse news content can be effective at reducing filter bubbles in news recommendations \cite{ookalkar2019pop}.

The need to combat filter bubble formation is clear \cite{resnick-cscw13}, but approaches vary. \citet{dong-www23} take an algorithmic approach and conduct simulations, defining bubbles as communities that have many inward but few outward connections in a bi-partite graph of users and items. They use a reinforcement learning algorithm to decide which community-connecting edges should be added to the graph to increase diversity.
Others, such as \citet{masrour2020bursting}, propose solutions based on algorithmic fairness criteria, while still others propose an attention-based modeling architecture to reduce the political homogenization effect in news recommendation \cite{shivaram2022reducing}.

In short, the extant research on this subject is temporally static, simulates user-controllable news recommender systems rather than examining the real world \cite{wang2022user}, lacks an appropriate interaction tool \cite{faridani2010, munson2013encouraging}, is overly descriptive \cite{harambam2019}, or focuses on user control in non-news (i.e., social media, movie/music recommendation, etc.) contexts \cite{bhargava2019gobo, taijala2018movieexplorer,jannach2016user}.

\subsection{Research Questions}
\label{sec.research_questions}

We conduct a user study of a political news recommendation system, where the {\em control} group has access to the traditional interaction mechanism of up-voting/down-voting a news article. The {\em treatment} group, however, has also access to an enhanced user interface (UI) (see Fig.~\ref{fig.transp.inter}) where they can adjust which political topics they are most interested in as well as their political preference for articles on each topic. We investigate the following research questions:

\begin{itemize}
    \item \textbf{RQ1}: How does a user's interaction with a political news recommender system affect the system's recommendation trajectory?
    \item \textbf{RQ2}: Do changes in the recommendation system's trajectory differ significantly for the control group versus the treatment group?
\end{itemize}

Both research questions are motivated by the extant literature in several ways. First, based on simulations of news recommender systems presented in \citet{liu2021interaction}, we know that users are increasingly presented more extreme, less diverse, and more homogeneous news articles. Yet, there are distinctions between ``challenge-averse'' and ``diversity-seeking'' individuals \cite{munson-chi10}, illustrating that personal characteristics and preferences affect the diversity of opinions to which people are exposed. Second, people are open to the possibility of manipulating the recommender system to increase exposure to diverse content \cite{harambam2019} --- and in fact are able to increase diversity through a transparency tool \cite{munson2013encouraging}. Greater diversity of content results in the perceived value of the recommendations initially going up but then going down \cite{ziegler2005}, i.e., the ability to self-navigate through online content does not increase its diversity \cite{faridani2010}.

For the {\em control} group, we expect that they will be presented increasingly more extreme and less diverse news over time, in line with \cite{liu2021interaction}.  Given that party affiliation is a key predictor of the online content with which people engage \cite{allen2022birds, tornberg2022}, that partisans are more entrenched in their beliefs \cite{Brewer2005} and thus more likely to be motivated by their preexisting beliefs \cite{kahan2015politically,Lodge2000}, and given the connections between partisanship and news extremeness \cite{tewksbury2015polarization, levendusky2016does}, we expect to see the largest shifts for users who consume more moderate news content. Regarding model accuracy, measured through up-vote ratio, we expect our models to improve as additional training data is fed into them, increasing the up-vote ratio over time.

Even though there is evidence in the literature for both ``challenge-averse'' users who prefer to see agreeable news and ``diversity-seeking'' users who are more amenable to diverse opinions \cite{munson-chi10}, we do not expect the simple mechanism of up-voting/down-voting articles to be sufficient for the diversity-seeking users to steer the system to less extreme and more diverse articles.

For the {\em treatment} group, we expect mixed results. We expect the interaction and transparency tool to enable ``challenge-averse'' individuals in the {\em treatment} group to steer the system to greater extremes than those in the {\em control} group, and enable those who are ``diversity-seeking'' in the {\em treatment} to steer it to less extreme and more diverse articles than those in the {\em control}. We expect mixed results for the up-vote ratio as well: the system should be able to learn user preferences better over time and hence lead to higher up-vote ratios; however, some users might use the transparency and interaction tool to drastically change the system and thus experience a lower up-vote ratio than those in the {\em control} group, as providing users with more control does not guarantee its effective utilization \cite{mitova2023}.

\section{Our Approach}
\label{sec.details}

\paragraph{Dataset} We focus on the political news domain in our user study, which not only has implications for policy making and electoral outcomes but is also likely to contain the type of ideologically polarizing content that can be most impactful --- and potentially harmful --- for society. 

We used the U.S.~political news dataset from \citet{liu2021interaction}, collected from September 2019 to August 2020. 
It contains articles from 41 news sources. Each article is annotated with a political stance rating in $\{-2, -1, 0, +1, +2\}$ by \textit{www.allsides.com}, where $-2$ indicates extreme liberal and $+2$ indicates extreme conservative. Each article is also annotated with one or more political topics.

We sampled 8,000 articles from each of the five political stances, which resulted in a total of 40,000 articles, summarized by topic in Table \ref{table: topical_distribution}. Given that an article can contain more than one topic label (e.g., ``immigration'' together with ``racism''), there are a total of 44,033 topic labels represented by the 40,000 articles. 5,000 articles were used to bootstrap the recommender system (explained in detail below), and the remaining 35,000 articles were retained as potential candidates for recommendation.

\begin{table}[t]
    \begin{subtable}[t]{.48\textwidth}
        \centering
        \begin{tabular}{@{\hskip 1pt}l@{\hskip 6pt}r@{\hskip 6pt}l@{\hskip 6pt}r}
            \toprule
            \textbf{Topic} & \textbf{\# articles} & \textbf{Topic} & \textbf{\# articles} \\
            \midrule
            abortion         & 1,988 & environment   & 2,854    \\ 
            foreign policy   & 5,759 & guns          & 2,781    \\   
            healthcare       & 5,999 & immigration   & 5,771    \\ 
            LGBTQIA          & 1,611 & racism        & 5,550    \\ 
            taxes            & 4,639 & trade         & 3,794    \\ 
            welfare          & 3,287 &    &   \\ 
            \midrule
            \# articles      & 40,000  & & \\ 
            \# labels        & 44,033  & & \\ 
            \bottomrule
        \end{tabular}
        \caption{}
        \label{table: topical_distribution}
    \end{subtable}
    \begin{subtable}[t]{.48\textwidth}
        \centering
        \begin{tabular}{@{\hskip 1pt}l@{\hskip 6pt}r@{\hskip 6pt}r@{\hskip 6pt}r}
        & & & \\
            \toprule
            &\textbf{Control} & \textbf{Treatment} & \textbf{Total} \\
            \midrule
            Male    & 26 & 30 & 56          \\
            Female  & 25 & 21 & 46           \\
            \midrule
            Democrat       & 17 & 18 & 35    \\
            Republican     & 23 & 24 & 47    \\
            Independent    & 10 & 9 & 19     \\
            \midrule
            Total  & 51 & 51 & 102\\
            \bottomrule
        \end{tabular}
        \caption{}
        \label{tab.user.stats}
    \end{subtable}        
    \caption{Topic distribution for articles used in the user study (a); and demographics of the user study participants (b).}    
\end{table}

\paragraph{User recruitment} We recruited participants from Amazon Mechanical Turk (AMT; \url{https://www.mturk.com/}) from September to December 2021. Users were required to reside in U.S., to have voted in the 2020 election, and to possess a sufficient level of political literacy as determined via a screening survey (see Appendix). Of the 850 users who completed this survey, 595 (70\%) answered at least two of the three qualification questions correctly and were subsequently invited to participate in the full recommendation system user study. Of the 595 invitees, 146 users participated in the final study, of which 44 were dropped from the study due to failed attention checks (see Appendix). The final study consists of 102 users divided randomly into control and treatment groups, the demographic information of which is presented in Table~\ref{tab.user.stats}.  

\paragraph{Recommender study pre-questionnaire}
\label{sec.pre} 
Before users were presented with news articles, we required them to complete a pre-questionnaire in which they revealed their ideological positions and personal interest in each of the 11 predominant political topics in the dataset (Table \ref{table: topical_distribution}). 

We built the pre-questionnaire based on the Pew surveys of U.S.~political typologies \cite{doherty2017political}, which asked how users agree or disagree (five-point Likert response) with statements about each of the topics in the political news dataset. For example, a user's response to the statement, ``Abortion should be legal in most cases,'' is used to estimate their  stance on abortion. Agreement or disagreement for all eleven topics is based on the statements presented in Table \ref{tab.pre_question}. 
A user's personal interest in each of the issues was determined by asking them about the extent to which they are interested in each of the eleven topics (five-point Likert response). This political stance and interest-related information was used for bootstrapping the recommendation algorithm.

\begin{table}[!t]
\small
\begin{center}
\begin{tabular}{ p{0.21\linewidth} | p{0.70\linewidth}}
\toprule
\multicolumn{2}{l}{\textbf{Profile Questionnaire}}\\
\midrule
abortion        & Abortion should be legal in most cases.                            \\ [4pt]
environment     & Stricter environmental regulations and laws are worth the costs.   \\ [4pt]
foreign policy  & Good diplomacy is the best way for the U.S. to ensure peace.       \\ [4pt]
guns            & Gun laws should be stricter than they are today.                   \\ [4pt]
healthcare      & Providing healthcare to Americans is the federal government’s responsibility. \\ [5pt]
immigration     & Immigrants strengthen the United States in many different ways.    \\ [4pt]
LGBTQIA         & Members of the LGBTIA+ community should have the right to marry.   \\ [4pt]
racism          & Changes are needed in American society to improve racial equality. \\ [4pt]
taxes           & The U.S. economic system unfairly favors powerful interests.       \\ [4pt]
trade           & U.S. involvement in the global economy is good for the country.    \\ [4pt]
welfare         & Poor people have hard lives because government programs do not do enough for them. \\ [1pt]
\bottomrule
\end{tabular}
\caption{Political stance pre-questionnaire. On a five-point Likert scale, users agreed or disagreed with each statement. \label{tab.pre_question}
}
\end{center}
\end{table}

\paragraph{Recommender system} We built a two-stage recommender algorithm. A personalized content-based recommender scores each potential news article. These scores were then adjusted based on the match of the user's political stance and interest in each topic to the political stance and topic of the candidate article.

\paragraph{Content-based recommender} We trained a personalized content-based recommender separately for each user. While recommendation systems are an active research area with many proposed deep learning solutions \cite{covington2016deep,ying2018graph}, our goal is not to develop a new recommendation algorithm but to generate a simpler system both to enhance interpretability and to avoid overfitting in the smaller data regime of the study.

We implemented a standard text-based recommendation system as follows. Each news article content was transformed into a tf-idf vector with a vocabulary size of $3,000$ words. A logistic regression classification model was trained to predict whether a user would like (up-vote) or dislike (down-vote) an article based on its content. After each user interaction (an up-vote or down-vote of an article), a stochastic gradient descent update was made to update the model and make new recommendations.

To address the ``cold-start'' challenge of recommender systems, we used users' responses to the pre-questionnaire (described above) as follows. We reserved $5,000$ of the $40,000$ articles for bootstrap purposes, and we bootstrapped the content-based recommender for each user based on their political stance and interest-related responses to the pre-questionnaire. For each user, we created a political stance vector $u_s$ with 11 entries, each of which corresponded to their stances on the 11 topics from the questionnaire. Then, for each topic, we sampled ``positive/up-vote'' articles that matched the user's stance, and ``negative/down-vote'' articles that were furthest away from the user's stance. For example, if a user's stance was $-2$ on abortion, the algorithm drew 25 articles from the abortion topic with a partisan score of {$-2$} as positive/up-vote, and then drew 25 articles from the same topic with a partisan score of $+2$ as negative/down-vote. Similar pairings were done for $-1$, $+1$, and $+2$. If the user's stance was $0$, however, the algorithm drew positive examples from stance $0$ and an equal number of negative examples from both {$+2$} and {$-2$}, assigning them sample weights of $0.5$. 
A logistic regression classifier was trained for each user using their personalized bootstrap dataset. 

\paragraph{Recalibration of the scores}
To operationalize the interaction mechanism presented in the enhanced UI (Figure \ref{fig.transp.inter}), we designed a two-component recommendation system that allows the users to directly influence the final ranking of the recommendations \cite{steck2018calibrated}. The first component is the content-based model described above. The second component captures how well the user's stance and interest in each topic (initially collected through the pre-questionnaire and then potentially modified through the interaction mechanism) match the stance and topic of the article. 

Let $u_s$ be a user's political stance vector, initialized using their answers to the political questions in the pre-questionnaire (Table \ref{tab.pre_question}). This vector has one entry per topic (i.e., 11 entries), each of which ranges from $-2$ to $+2$. Let $u_t$ be user's interest vector on each topic (11 entries), indicating their interest in each topic, initialized based on their pre-questionnaire. Similarly, let $a_s$ be an article's political stance on each topic,
and let $a_t$ be the article's binary vector indicating its topic(s). 

Let the recommendation score of the content-based recommender for an article be $s_r$ (i.e., the probability of the ``up-vote'' class in the binary classifier). The final recommendation score for that article combines $s_r$ with the match between the user's interest and political stance vectors and those of the article:
\begin{equation}
\label{eq.calibrate}
{s} = \lambda s_r + (1-\lambda) (Sim(u_t, a_t) + Sim(u_s, a_s)) / 2
\end{equation}
where $Sim$ is the cosine similarity of the two vectors. 
The transparency and interaction tool (Figure \ref{fig.transp.inter}) allowed users to adjust the average political stance and the proportion of the top-ranked articles, which was implemented algorithmically by adjusting $u_t$ and $u_s$ vectors. We set  $\lambda = 0.4$ in our experiments, as our preliminary analysis showed that it provides the best balance between the content-based classifier preference and the user's stance and interest profile. 

When users were presented with an article recommended by the system, they could up-vote the article, down-vote the article, or skip it. If the users up-voted or down-voted, the article was added to the training set, the recommender was retrained using stochastic gradient descent, and a new recommendation was presented on the next appearing page. Skipped articles generated new recommendations using the existing system. Users were presented with one article at a time until they up/down-voted a total of $30$ articles.

\paragraph{Transparency and interaction}
The transparency and interaction tool (Figure \ref{fig.transp.inter}) in the treatment group exposed users to information about the recommender system, allowing them to adjust the recommender system's ranking for each topic via political stance and interest sliders. This interface reflected the current state of the recommender system for that user via the average political stance and the topic distribution of the top $K$ ranked articles ($K=200$ in our experiments below). The left panel of Figure \ref{fig.transp.inter} (``Political Stance'') showed the average political stance of the top-ranked articles on each topic; the right panel (``Interest'') reflected the proportions of the topics in the top-ranked articles. Any change to a slider's position was recorded immediately by updating $u_t$ and $u_s$ in Eq.\ref{eq.calibrate} through a binary search (trying possible $u_t$ and $u_s$ values and re-ranking the news articles) to quickly find the new $u_t$ and $u_s$ values reflecting the desired topic interest and political stance among the top-ranked articles. The recommender system, the top-ranked articles, and the locations of the sliders were then updated.
For example, if dissatisfied with the system's recommendations, the user reflected in Figure \ref{fig.transp.inter} could use the political stance slider to move towards the center on the abortion topic or to the right on taxes. With the interest slider, the same process occurs to affect the proportion of articles presented on each topic. Users also had the option of resetting sliders back to the last system recommendation update.

Access to the transparency and interaction tool was provided to users in the treatment group via a link at the top of each recommendation page. They could thus opt to view their system-based profile at any time; yet, to account for users who might not use the link, users were also automatically directed to this page after reading and scoring every five articles. To account for users who were overly-dissatisfied with the content they were reading, they would be automatically directed to the transparency and interaction tool page after three consecutive down-vote actions. Instructions regarding how the user could adjust the system and details about what the sliders' positions meant were provided at the top of the interaction page. Users could spend as much time as they liked on this page before continuing to the next news article in the recommendation process.

\paragraph{Recommender study post-questionnaire}
Given that users' perceived usefulness and perceived ease of use of a particular information technology is essential for their adoption of such technology \cite{davis1989, venkatesh2003, venkatesh2008technology}, after users up-voted/down-voted a total of 30 articles (approximately a one-hour session), they were then presented with a post-questionnaire to gauge their perceptions about filter bubbles and the recommender system. Users in both the control and treatment groups answered questions (five-point Likert response) regarding the extent to which they enjoyed the system (\emph{Qa}), the extent to which they were presented with diverse articles (\emph{Qb}), and the extent to which the study helped them learn more about how news recommender systems work (\emph{Qd}). This single-question approach is consistent with research that bases users' opinions of recommender systems based on a single question \cite{faridani2010, tsai2020}.

While the control group lacked access to the transparency tool, we could assess whether their initial responses (\emph{Qb}) remained unchanged after being presented with information about the articles they were shown. After answering \emph{Qb}, users in the control group were presented with a histogram of the political stances of the articles they were shown (Figure \ref{fig.transparency} in the Appendix shows an example). Users in the control group responded once again to the diversity-of-news question (\emph{Qc}), allowing us to use the difference between \emph{Qb} and \emph{Qc} as a measure of how clarity about the news the user consumed impacted their beliefs about the diversity of news, particularly for users having no access to the interaction tool.\footnote{Code and user survey and interaction data are available at: \url{https://github.com/IIT-ML/icwsm-2024-filter-bubbles}.}

\paragraph{Measures}
Inspired in part by prior work on diversity in recommendation systems~\cite{castells2021novelty}, we identified four key measures to evaluate the system and compare the control and treatment groups. 

\noindent \textit{(1) Average political stance:} This measure captures the overall ideological stance of the recommender system for user $u$ for a given time period. The recommender system political stance score for user $u$ is the average of the political stance of the top $K$-ranked articles from the recommender system. Let $a_j^u$ be the article ranked at position $j$ for user $u$ and $s(a)$ be the political stance of the article $a$. The recommender system political stance score is:
\begin{equation}
\textit{Political Stance}(u) = \frac{1}{K}\sum\limits_{j=1}^{K} s(a_j^u)
\label{eq.stance}
\end{equation}

\noindent \textit{(2) Average extremeness:} This measure captures the overall extremeness of the recommender system for user $u$ for a given time period. Because political stance ranges from $-2$ (extreme liberal) to $+2$ (extreme conservative), where $0$ represents neutral, we use the absolute value of political stance for the extremeness measure. Hence, the recommender system extremeness score for user $u$ is the average of the absolute value of the political stance of the top $K$ ranked articles by the recommender system. The recommender system extremeness score is:
\begin{equation}
\mathit{Extremeness}(u) = \frac{1}{K} \sum\limits_{j=1}^{K} |s(a_j^u)|
\label{eq.exreme}
\end{equation}

\noindent \textit{(3) Diversity:} This measure captures the diversity of political stances of the recommendations. Let $p_l^u$ be the proportion of articles having political stance $l$ in the top $K$ recommendations for user $u$. We measure diversity using the normalized political stance entropy as follows:
\begin{equation}
\mathit{Diversity}(u) = \frac{\sum\limits_{l=-2}^{+2} -p_l\times \log(p_l^u)}{\log(5)}
\label{eq.diversity}
\end{equation}
where $\log(5)$ is a normalization constant to ensure the entropy of a 5-category distribution is between 0 and 1, with 1 representing maximum diversity.

\noindent \textit{(4) Up-vote ratio:} Our measure of system accuracy is the proportion of the recommended articles liked by the user. Let $r_i^u$ be $1$ if the user $u$ up-voted the $i^\textrm{th}$ article shown to them, and $0$ otherwise. \textit{Up-vote Ratio} is defined as:
\begin{equation}
\textit{Up-vote Ratio}(u) = \frac{1}{N}\sum_{i=1}^N r_i^u
\label{eq.uvr}
\end{equation}

\section{Evaluation Methodology and Findings}
These four measures --- political stance, extremeness, diversity, and up-vote ratio --- are given our entire focus as we answer the research questions presented in \S\ref{sec.research_questions}. 

RQ1 --- \textit{How does a user's interaction with a political news recommender system affect the system's recommendation trajectory?} --- considers whether users are presented with progressively more extreme/moderate, liberal/center/conservative, diverse/homogeneous, and enjoyable/less-desirable articles while interacting with the system. We answer RQ1 twice, once for the control group and once for the treatment group.

We first quantify the \emph{initial} extremeness, political stance, and diversity states of the recommender system for a user (the \texttt{begin} value) using the top $K$-ranked articles for that user prior to the first article being shown to the user.\footnote{We used $K=200$ in our experiments to give all topic and political stance combinations (11 topics, 5 political stances = $11\times5=55$ possibilities) a reasonable chance to be included in the top $K$ articles.} This initial top-$K$ ranking is based on the model bootstrapped from the pre-questionnaire responses (see Table \ref{tab.pre_question}), i.e., the user's political stance and interest for the 11 topics. The \emph{final} measures (\texttt{end}) are similarly computed after the last article has been presented. Whereas extremeness, political stance, and diversity are captured at a large-scale using the top-$K$-ranked articles for the user, the up-vote ratio is based on the up/down votes the user assigned to each of the presented articles, and hence the \texttt{begin} up-vote ratio is captured via the first ten articles that the user up/down-voted, and the \texttt{end} value is computed based on the last ten articles. More formally, for measure $m$ let $m_b$ be the \texttt{begin} value and $m_e$ be the \texttt{end} value. For users in the control group and treatment group, separately, RQ1 considers whether $\delta_m=m_e-m_b$ is significantly different from zero.

In addition to these main effects, we also expect there to be heterogeneous effects based on user attributes. For example, \citet{munson2013encouraging} find that conservative and liberal users respond differently to a recommendation interface that shows users summaries of the partisan lean of their reading habits. In order to operationalize this, we consider the initial state of the recommendation system, which, recall, is seeded based on the user survey. To balance our ability to identify heterogeneous effects with ensuring a sufficient sample size, for each measure we assign users into one of three bins, based on the initial state of the system for that user. That is, for each measure, we rank all 102 users based on their \texttt{begin} values $m_b$. We then assign each user to \texttt{low}, \texttt{medium}, and \texttt{high} subgroups based on their rank,\footnote{Each subgroup had $102/3=34$ users, but they do not always divide evenly into control and treatment groups. See \S\ref{sec.discussion} for a discussion of sample sizes.} and let $g \in$ \texttt{\{all, low, medium, high\}}. For each measure, $m$, we compute $\delta_m^g=m_e^g-m_b^g$ and test for whether it is significantly different from zero.

RQ2 --- \textit{Do changes in the recommendation system's trajectory differ significantly for the control group versus the treatment group?} --- employs a between-group comparison to examine whether the \texttt{begin} to \texttt{end} differences are larger/smaller for users assigned to the control group relative to users assigned to the treatment group. More formally, let $\Delta_m^g=\delta_m^g(\mathrm{Treatment}) - \delta_m^g(\mathrm{Control})$. We test for whether $\Delta_m^g$ is significantly larger/smaller than zero. Such tests are conducted in the context of the \texttt{all} group as well as the \texttt{low}, \texttt{medium}, and \texttt{high} subgroups.

\subsection{Findings}
\label{sec.findings}

\begin{figure*}[!ht]
    \centering
    \subfloat[Extremeness\label{fig.m.e}]{\includegraphics[width=.99\linewidth]{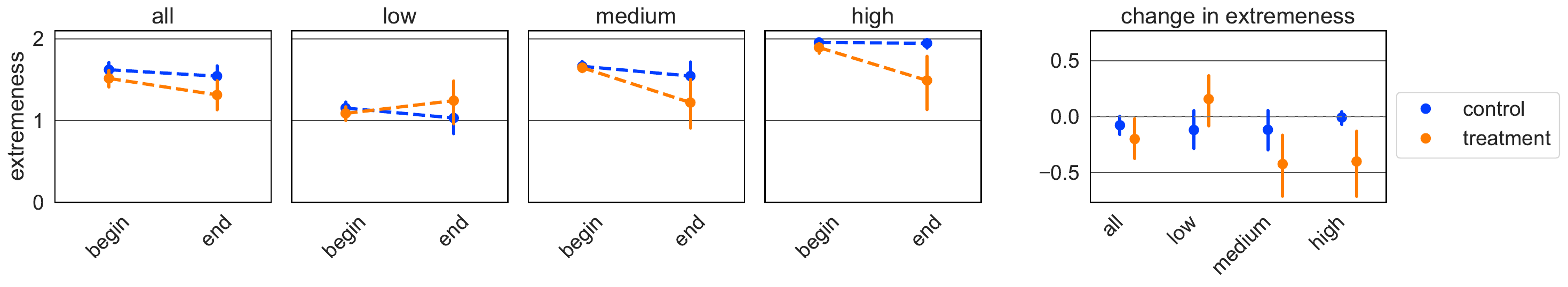}}\vspace{3pt} \\
    \subfloat[Political Stance\label{fig.m.s}]{\includegraphics[width=.99\linewidth]{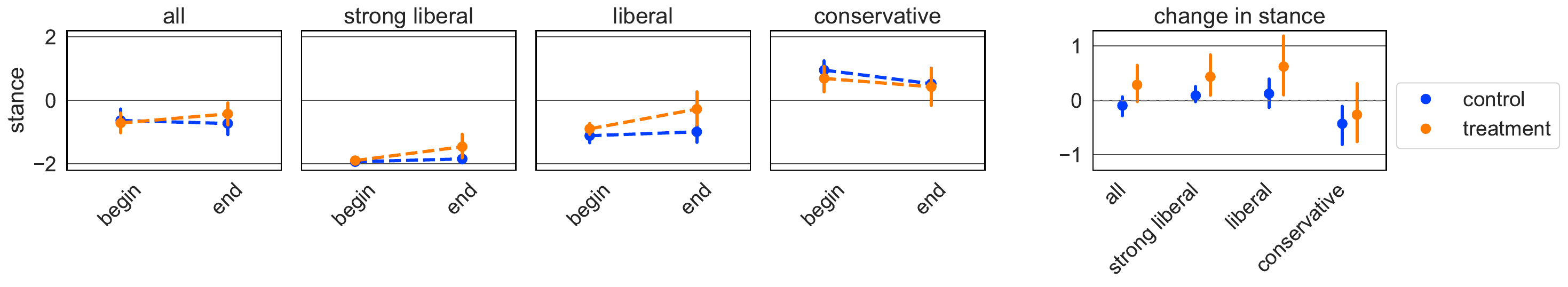}}\vspace{3pt}\\
    \subfloat[Diversity\label{fig.m.d}]{\includegraphics[width=.99\linewidth]{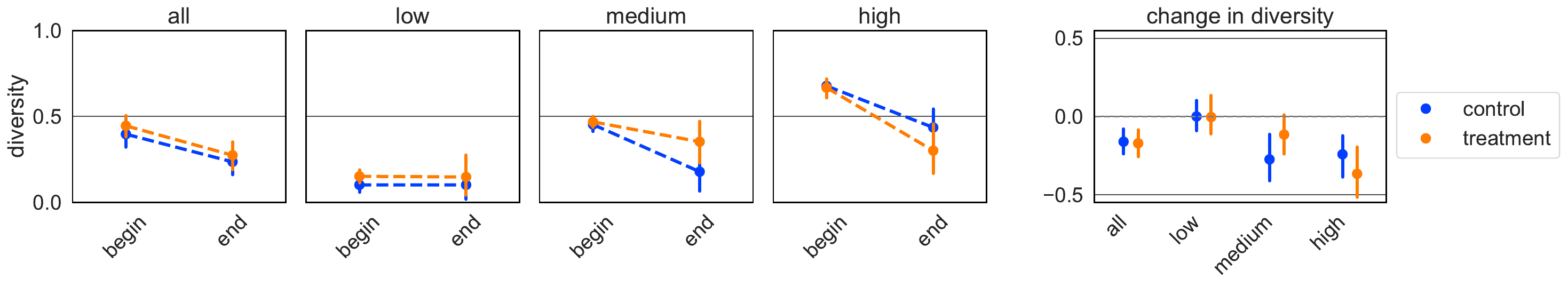}}\vspace{3pt}\\
    \subfloat[Up-vote Ratio \label{fig.m.u}]{\includegraphics[width=.99\linewidth]{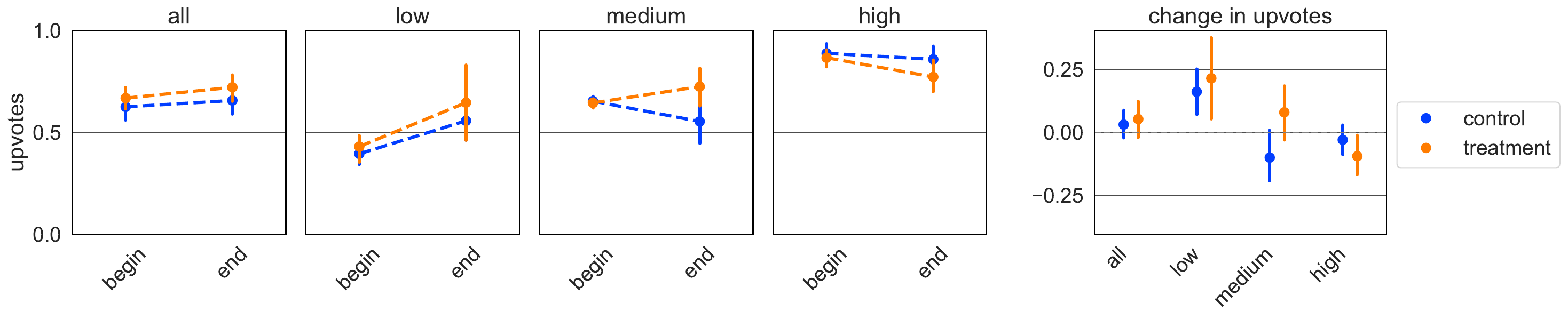}}\\
    \caption{Beginning ($m_b^g$) and ending ($m_e^g$) values, with bootstrapped 95\% confidence intervals (1000 bootstrap samples), for each measure and subgroup, where subgroups are determined by the initial value of the system for each user. Rightmost graphs present changes over time (i.e., $\delta_m^g=m_e^g-m_b^g)$.}
    \label{fig.measures}
\end{figure*}

Figure \ref{fig.measures} presents the results for each measure. The first graph in each row shows the \texttt{begin} ($m_b$) and \texttt{end} ($m_e$) values for \texttt{all} users within the control and treatment groups. The second, third, and fourth graphs in each row present the same information based on, respectively, the \texttt{low}, \texttt{medium}, and \texttt{high} subgroups. The final, rightmost graph in each row plots the changes in these data for each group ($\delta_m^g$).
The first four columns of Figure \ref{fig.measures} are primarily used to answer RQ1, while the rightmost column of Figure \ref{fig.measures} is primarily used to answer RQ2. Error bars in each plot are bootstrapped 95\% confidence intervals (1000 bootstrap samples).

Table \ref{tab.pvalues} presents t-tests to assess within-group comparisons (i.e., whether $m_e^g$ and $m_b^g$ are significantly different) as well as between-group comparisons (i.e., whether $\delta_m^g$ differs significantly between the control and treatment groups). A response to RQ1 is provided in the ``Begin vs End'' columns of each measure (1, 2, 4, 5, 7, 8, 10, 11), and $p$ values are computed using two-tailed paired t-tests based on a comparison of the values of $m_b^g$ and $m_e^g$.\footnote{Two-tailed t-tests examine whether the \texttt{begin} and \texttt{end} values are significantly different from each other; paired are used because these are the \texttt{begin} and \texttt{end} values for the same users.} A response to RQ2 is provided in the ``Change'' column of each measure (3, 6, 9, 12), where $p$ values are computed using one-tailed non-paired t-tests.\footnote{One-tailed t-tests examine whether the difference is significantly larger/smaller for the treatment group; non-paired are used because these are different users (i.e., control versus treatment).} Each measure is considered in turn below.

\paragraph{Extremeness}
Based on \citet{liu2021interaction}, we expect extremeness to go up for the control group, as the feedback loops to exacerbate the filter bubbles and the simple interaction mechanism of up-voting/down-voting will be inadequate for course correction. For the treatment group, we expect mixed results, where the interaction tool to satisfy the requests of both challenge-averse and diversity-seeking groups \cite{munson-chi10}.

RQ1: Within-group comparisons in Figure \ref{fig.m.e} show that extremeness decreased for the entire sample (\texttt{all}) for both the treatment ($\delta=-.20, p=.025$) and control groups ($\delta=-.08$, $p=.065$).
For the \texttt{low}, \texttt{medium}, and \texttt{high} subgroups, the within-group comparison results are mixed. For users starting with either \texttt{medium} or \texttt{high} extremeness, those in the treatment group generated significant decreases in extremeness ($\delta=-.43, p=.005$ and $\delta=-.40,p=.041$, respectively), while the those in the control group experienced very little change
. For the \texttt{low} subgroup, however, extremeness increased slightly for those in the treatment group ($\delta=.15, p=.196$), while it decreased slightly in the control group ($\delta=-.12, p=.200$). In sum, for those in the treatment group, extremeness decreased at statistically significant levels for the \texttt{all} group and the \texttt{medium} and \texttt{high} subgroups. For the control group, changes from beginning to end were insignificant.

\begin{table*}[t]\centering
\footnotesize
\begin{tabular}{lccc|ccc|ccc|ccc}
&\multicolumn{3}{c}{\textbf{Extremeness}} &\multicolumn{3}{c}{\textbf{Political Stance}} &\multicolumn{3}{c}{\textbf{Diversity}} &\multicolumn{3}{c}{\textbf{Up-vote Ratio}} \\\cmidrule{2-13}
&\multicolumn{2}{c}{\textbf{Begin vs~End}} &\multicolumn{1}{c}{\textbf{Change}} &\multicolumn{2}{c}{\textbf{Begin vs~End}} &\multicolumn{1}{c}{\textbf{Change}} &\multicolumn{2}{c}{\textbf{Begin vs~End}} &\multicolumn{1}{c}{\textbf{Change}} &\multicolumn{2}{c}{\textbf{Begin vs~End}} &\textbf{Change} \\\cmidrule{2-13}
&\textbf{C~vs~C} &\textbf{T~vs~T} &\multicolumn{1}{c}{\textbf{C~vs~T}} &\textbf{C~vs~C} &\textbf{T~vs~T} &\multicolumn{1}{c}{\textbf{C~vs~T}} &\textbf{C~vs~C} &\textbf{T~vs~T} &\multicolumn{1}{c}{\textbf{C~vs~T}} &\textbf{C~vs~C} &\textbf{T~vs~T} &\textbf{C~vs~T} \\\cmidrule{2-13}
&\textbf{(1)} &\textbf{(2)} &\multicolumn{1}{c}{\textbf{(3)}} &\textbf{(4)} &\textbf{(5)} &\multicolumn{1}{c}{\textbf{(6)}} &\textbf{(7)} &\textbf{(8)} &\multicolumn{1}{c}{\textbf{(9)}} &\textbf{(10)} &\textbf{(11)} &\textbf{(12)} \\\midrule
\textbf{All} &.065 &\textbf{.025} &.103 & .296 & .084 &\textbf{.022} &\textbf{\underline{.000}} &\textbf{\underline{.001}} &.435 & .292 & .157 & .324 \\
\textbf{Low} & .200 & .196 &\textbf{.034} & .245 &\textbf{.050} & .064 & .999 &.954 & .482 &\textbf{\underline{.003}} &\textbf{.045} & .416 \\
\textbf{Medium} & .209 &\textbf{\underline{.005}} &\textbf{.023} &.381 &\textbf{.036} &.055 &\textbf{\underline{.002}} &.091 &\textbf{.047} &.302 &.118 &\textbf{.033} \\
\textbf{High} &.774 &\textbf{.041} &\textbf{.024} &\textbf{.035} &.225 &.416 &\textbf{\underline{.003}} &\textbf{\underline{.000}} &.091 &.083 &\textbf{.049} &.220 \\
\bottomrule
\end{tabular}
\caption{$p$-values from $t$-tests of significance, $p \le .05$ in bold. $p$-values smaller than $0.0125$ (i.e., accounting for the Bonferroni correction for testing four hypotheses per measure) are in bold and underlined. Rows represent the aggregated group (``All'') and all subgroups (\textit{Political Stance} subgroups correspond with ``strong liberal,'' ``liberal,'' and ``conservative'').}\label{tab.pvalues}
\end{table*}

RQ2: We observe between-group comparisons in the rightmost graph of Figure \ref{fig.m.e} and column 3 of Table \ref{tab.pvalues}, showing that the change in extremeness was consistently larger for the treatment group relative to the control group. These differences are statistically significant for the \texttt{low}, \texttt{medium}, and \texttt{high} subgroups ($\Delta=.28, p=.034$, $\Delta=-.31, p=.023$, and $ \Delta=-0.39, p=.024$, respectively).  That is, when extremeness decreased for the treatment group (\texttt{medium} and \texttt{high} subgroups), it decreased significantly more than the control group. When extremeness increased for the treatment group (\texttt{low} subgroup), it increased significantly more than the control group.

These results point to heterogeneous effects of the proposed system --- empowered with the transparency and interaction tool, the treatment group was able to make bigger and more significant changes to the system. However, this did not result in reducing extremeness for everyone: users whose system was initially \texttt{medium} or \texttt{high} reduced extremeness, while users whose system was initially \texttt{low} increased extremeness.

\paragraph{Political Stance}
Our expectations for changes in political stance are the same as those for extremeness: the control group solidifies their initial political assignments, and the treatment group is able to make bigger changes (positive or negative). Figure \ref{fig.m.s} shows the average political stance of the top $K$-ranked articles for each subgroup. To provide a more intuitive explanation of this measure, we swap the \texttt{low}, \texttt{medium}, and \texttt{high} subgroup labels with their respective ideologies. Given that the average \texttt{begin} values for these subgroups were, respectively, $-1.9$, $-1.0$, and $0.9$, we label the three subgroups ``strong liberal,'' ``liberal,'' and ``conservative.''\footnote{Even though more people self-identified as Republican (47 users) than Democrat (35 users) in the pre-questionnaire (Table \ref{tab.user.stats}), the initial state of the recommender was slanted slightly liberal. Recall that this initial state was based on user-provided responses to select Pew survey questions (Table \ref{tab.pre_question}); perhaps Republican users in our sample held more liberal views, or perhaps political ideology had shifted leftward since the Pew survey was published in 2017.} 

RQ1: Considering the entire sample (\texttt{all}), we observe that the political stance of users' news content shifted slightly towards the center for the treatment group ($\delta=.29, p=.084$) but changed little for the control group ($\delta=-.10, p=.296$). By subgroup, the most noticeable changes among users in the treatment group are for strong liberals ($\delta=.44, p=.050$) and liberals ($\delta=.62, p=.036$), who consumed significantly less partisan articles. For the control group, the most noticeable change is for conservative users, who also consumed less partisan articles ($\delta=-.43, p=.035$). These results are largely consistent with those of the extremeness measure: users generally moved the recommender towards less partisan news content.

RQ2: For the between-group comparisons (rightmost graph of Figure \ref{fig.m.s} and column 6 of Table \ref{tab.pvalues}), we observe that the treatment group experienced larger shifts towards the center than the control group when considering the \texttt{all} group ($\Delta=.38, p=.022$), as well as for strong liberals ($\Delta=.35, p=.064$) and liberals ($\Delta=.50, p=.055$). For conservatives, those in both the control and treatment groups directed the system towards the center, but the between-group difference was not statistically significant.

\paragraph{Diversity} 

We expect the system to take the control group to less diverse articles \cite{liu2021interaction};
For the treatment group, we expect the interaction tool to either increase or decrease diversity depending on users' preexisting characteristics \cite{munson-chi10}.

While extremeness and average political stance are certainly informative, two users with the same extremeness score might view very different types of news. For example, one user with an extremeness score of 1 might have read a diverse set of articles (e.g., from both $+1$ and $-1$ sources), while another user with an extremeness score of 1 might have viewed articles representing only a single political stance (e.g., only $-1$). To assess this, we analyze the political stance diversity of the top $K$-ranked articles, measured through normalized entropy of the political stance distribution of the articles (Eq.~\ref{eq.diversity}). 

RQ1: Figure \ref{fig.m.d} shows that, for the (\texttt{all}) group, diversity decreased significantly for both the control ($\delta=-.16, p<.001$) and treatment ($\delta=-.17, p=.001$) groups (c.f., Table \ref{tab.pvalues}, columns 7 and 8). This was not unexpected given the tendency for filter bubble-like conditions to arise. Specifically, users may find that the initial personalization of results by the system are excessively broad and not reflective of users' self-perceptions.

Considering subgroups separately, the most noticeable changes among the treatment group was the \texttt{high} subgroup, which decreased significantly in diversity ($\delta=-.37, p<.001$), and to a lesser extent the \texttt{medium} subgroup, which decreased slightly ($\delta=-.12, p=.091$). For the control group, users in both the \texttt{medium} ($\delta=-.27, p=.002$) and \texttt{high} ($\delta=-.24, p=.003$) subgroups showed significant decreases in diversity. There was no discernible change in diversity of news content for users in the \texttt{low} subgroup for either the control 
or treatment groups.

RQ2: Regarding between-group comparisons (rightmost graph of Figure \ref{fig.m.d} and column 9 of Table \ref{tab.pvalues}), there are no noticeable differences between the treatment and control groups when considering the diversity of news viewed by \texttt{all} users. However, for those in the \texttt{medium} subgroup, the control group exhibited a sharper decrease in diversity than the treatment group ($\Delta=.15, p=.047$). We suspect that user feedback led to homogenization of the content recommender in the control group, while the enhanced UI available to users in the treatment group allowed them to maintain higher diversity.

\paragraph{Up-vote Ratio}

Based on the large body of research in machine learning and recommender systems, we expect the up-vote ratio to increase for users in both the control and treatment groups, as the system collects more training data and learn the user preferences better. However, for the treatment group, users can effect the state of the recommender directly, and hence can lead to unexpected results if the users change the system drastically by significantly overpowering the underlying content-based recommender system.

RQ1: Regarding within-group comparisons of the up-vote ratio, presented in Figure \ref{fig.m.u} and columns 10 and 11 of Table \ref{tab.pvalues}, we observe for \texttt{all} users only a slight increase in up-vote ratios for both the control
and treatment
groups. However, when considering subgroups separately, we observe significant increases for the \texttt{low} subgroup for both control ($\delta=.17, p=.003$) and treatment ($\delta=.19, p=.045$) groups. Conversely, for the \texttt{high} subgroup, the treatment group exhibited a significant decrease in its up-vote ratio ($\delta=-.09, p=.049$); the control group also exhibited a decrease to a lesser extent ($\delta=-.06, p=.083$). We will revisit this finding when considering the results from the post-questionnaire analysis.

RQ2: For the between-group comparisons (rightmost graph of Figure \ref{fig.m.u} and column 12 of Table \ref{tab.pvalues}), the most notable difference between control and treatment groups is for the \texttt{medium} subgroup ($\Delta=.14, p=.033$), where up-vote ratios increased for the treatment group but decreased for the control group. 

\paragraph{Summary of results}
Our analysis showed that the proposed interaction mechanism allowed users to exert greater control over the recommended articles, albeit in varying ways. The treatment group  reduced \emph{extremeness} significantly when the initial state was \texttt{medium} or \texttt{high} extreme, whereas the control group did not shift significantly.
Yet, for \emph{political stance}, all subgroups moved towards the center for both the control and treatment groups. Reductions in extremeness --- both formally and in terms of political stance --- may have been reduced, but this was accompanied with a decrease in the \emph{diversity} of users' news content. The results was also mixed for the \emph{up-vote ratio}: users who were initially less satisfied with news content (\texttt{low} up-vote ratio) could significantly improve their news consumption experience; those initially more satisfied (\texttt{high} up-vote ratio) became more dissatisfied.

\begin{figure}[t]
    \centering
    \includegraphics[trim = {5cm 1cm 5cm 1cm}, width=0.8\linewidth]{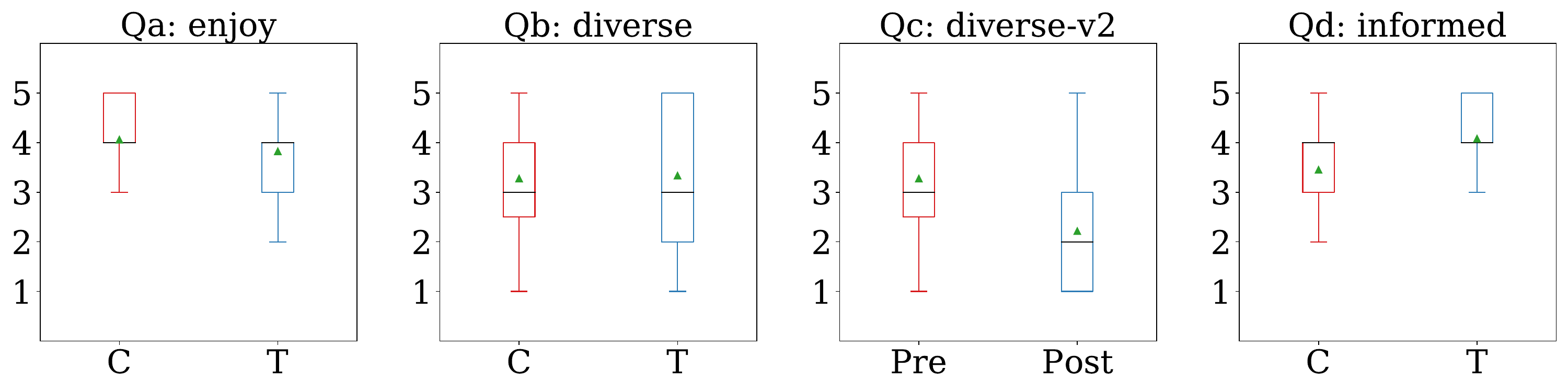} \hfill
    \caption{Post-questionnaire results. 1: strongly disagree, 5: strongly agree. Triangles represent means.}
    \label{fig.post_Qs}
\end{figure}

\paragraph{Post-questionnaire analysis}
 Figure \ref{fig.post_Qs} presents the post-questionnaire results. We find that users in the treatment group, though tending to have higher up-vote ratios in general, expressed less preference for reading articles based on the presentation of the recommender system (\emph{Qa: enjoy}). 
 It is possible that the added cognitive load of the enhanced interface hampered the user experience, bu it is also possible that it encouraged users to explore articles that were less aligned with their stances and interests, leading to unpleasant cognitive dissonance, particularly for those experiencing moderate enjoyment, i.e., the \texttt{medium} subgroup in Figure \ref{fig.m.u}. Future work should integrate composite measures of enjoyment as well as conduct extensive follow-up interviews to understand the reasoning behind these differences.

 There were no differences between groups in terms of the perceived level of exposure to the political diversity of the news articles to which users were presented (\emph{Qb: diverse}); users from both the control and treatment groups agreed that they had been presented with diverse political perspectives. However, after answering \emph{Qb}, when individuals in the control group were presented with the political stance distribution of the articles they were actually shown (see Figure \ref{fig.transparency} in Appendix as an example), and when they were asked a second time whether they thought they had been exposed to a diverse set of news articles, we observe a substantial drop in agreement. Presented in \emph{Qc: diverse-v2}, the first response (``pre'') is significantly higher than the second response (``post''). This indicates that when provided with transparency about the nature of the recommender system, users from the control group eventually realized that they were presented with less diverse articles than they thought. 
 Finally, users in the treatment believed much more strongly that, having participated in this study, they were more informed about how news recommender systems work (\emph{Qd: informed}). Taken together, the post-questionnaire results indicate that the enhanced UI keeps the user better informed, but it may come at a cost of greater UI complexity.

\section{Discussion, Social Impacts, Limitations}
\label{sec.discussion}
Our efforts to understand user engagement and news content delivery in a real-world setting have shown 
that, 
with the right information and interaction tools, people can counter filter bubbles in some aspects (extremeness) while reinforcing them in other ways (diversity). 
If user interaction tools were made available at scale, one might observe a decrease in polarizing content. This would facilitate opportunities for discourse among politically disparate groups --- conditions fundamental for healthy democratic institutions and for the development of policies reflecting the public's opinions and preferences \cite{habermas1989, lewandowsky2012}.

Of course, many other factors are at play here, including trust in the media \cite{guess2021consequences}, personal experiences \cite{druckman2021affective}, and online ideological segregation \cite{bail2021breakingprism, iandoli2021metaanalysis, mosleh2021partisanties}. Users' political preferences are particularly important, as existing research shows that individuals, especially those with firm beliefs, prefer to receive ideologically consistent \cite{
stier2020populist, ekstrom2023}, sometimes extreme content \cite{zhang2023}. This has prompted research focusing on how to ``nudge'' people out of their filter bubbles.\footnote{See, for example, \citet{srba2023, masrour2020bursting,pennycook2021shiftingattention}, 
although the evidence is mixed on whether these methods help \cite{aslett2022} or foster a ``boomerang effect'' \cite{casas2023exposure}.}

In terms of limitations to our work, as in any controlled study, the behavior of users in a short-horizon study may differ somewhat from their long-term use of commercial recommendation systems. 
Furthermore, while our results are based on $\sim$100 total hours of real system interactions from 102 users in a controlled environment, future work that considers larger sample sizes in natural environments would enhance the external validity of this study and also enable a closer investigation of how filter bubble behaviors vary by topic. To explore the impact of the small sample size on the main results, a post-hoc power analysis revealed that our significant findings in Table~\ref{tab.pvalues} have moderate to high power ($\sim$.7-.99). However, some of the comparisons not found to be statistically significant may be in part influenced by these results' lower power.

Finally, we have in many ways provided users with a more nuanced control interface (Figure \ref{fig.transp.inter}) than prior work, where the users are able to express a variety of political lean and interest across 11 topics (for e.g., liberal on abortion with low interest, conservative on taxes with high interest, etc.), rather than a single political identity (e.g., Democrat, Republican). While this implementation can allow the users to have diversity across topics, it may also lead to a lack of diversity on a given topic. A more comprehensive approach could be implemented where users can express, for example, that they would like to see only liberal views on abortion while simultaneously preferring both conservative \emph{and} liberal content on taxes. Such an approach would of course make the interface much more complicated and potentially overwhelming for users, and future work can determine whether the added complexity of such interfaces would be justified.

\section{Conclusions}
\label{sec.conclusions}

News recommendation systems can affect civic discourse. Thus, any variants in such systems have the potential to exacerbate hyperpartisanship and misinformation. The goal of this study was to understand the effect that transparency and interaction mechanisms have on these systems. Our results suggest that giving users greater control over news recommendation systems can substantially change the type of news they see. Users who are initially shown extreme content can more efficiently move towards less extreme content if desired; likewise, users who are initially shown less extreme content can move towards more extreme content if desired. We find heterogeneous effects based on the initial state of the recommender system, and we call for future work that investigates further how user attributes interact with new interaction mechanisms for recommendation systems.

\section*{Acknowledgements}
This work was supported by the NSF Award \#1927407. AC was funded in part by the Tulane's Jurist Center for Artificial Intelligence and by Tulane's Center for Community-Engaged Artificial Intelligence.

\bibliography{main}

\section{Ethics Checklist}

\begin{enumerate}

\item For most authors...
\begin{enumerate}
    \item  Would answering this research question advance science without violating social contracts, such as violating privacy norms, perpetuating unfair profiling, exacerbating the socio-economic divide, or implying disrespect to societies or cultures?
    \answerYes{Yes, the study conducted was granted IRB approval by the authors' institutions.}
  \item Do your main claims in the abstract and introduction accurately reflect the paper's contributions and scope?
    \answerYes{Yes.}
   \item Do you clarify how the proposed methodological approach is appropriate for the claims made? 
    \answerYes{Yes, see ``Evaluation Methodology and Findings'' section.}
   \item Do you clarify what are possible artifacts in the data used, given population-specific distributions?
    \answerYes{Yes, see ``Our Approach'' section.}
  \item Did you describe the limitations of your work?
    \answerYes{Yes, see ``Discussion, Social Impacts, Limitations'' section.}
  \item Did you discuss any potential negative societal impacts of your work?
    \answerYes{Yes, see ``Discussion, Social Impacts, Limitations'' section.}
      \item Did you discuss any potential misuse of your work?
    \answerYes{Yes, see ``Discussion, Social Impacts, Limitations'' section.}
    \item Did you describe steps taken to prevent or mitigate potential negative outcomes of the research, such as data and model documentation, data anonymization, responsible release, access control, and the reproducibility of findings?
    \answerYes{Yes, see ``Our Approach'' section.}
  \item Have you read the ethics review guidelines and ensured that your paper conforms to them?
    \answerYes{Yes.}
\end{enumerate}

\item Additionally, if your study involves hypotheses testing...
\begin{enumerate}
  \item Did you clearly state the assumptions underlying all theoretical results?
    \answerYes{Yes, see ``Our Approach'' and ``Evaluation Methodology and Findings'' sections.}
  \item Have you provided justifications for all theoretical results?
    \answerYes{Yes, see ``Related Work'' and ``Our Approach'' sections.}
  \item Did you discuss competing hypotheses or theories that might challenge or complement your theoretical results?
    \answerYes{Yes, see ``Related Work'' section.}
  \item Have you considered alternative mechanisms or explanations that might account for the same outcomes observed in your study?
    \answerYes{Yes, see ``Discussion, Social Impacts, Limitations'' section.}
  \item Did you address potential biases or limitations in your theoretical framework?
    \answerYes{Yes, see ``Our Approach'' section.}
  \item Have you related your theoretical results to the existing literature in social science?
    \answerYes{Yes, see ``Introduction'' and ``Related Work'' sections.}
  \item Did you discuss the implications of your theoretical results for policy, practice, or further research in the social science domain?
    \answerYes{Yes, see ``Discussion, Social Impacts, Limitations'' section.}
\end{enumerate}

\item Additionally, if you are including theoretical proofs...
\begin{enumerate}
  \item Did you state the full set of assumptions of all theoretical results?
    \answerNA{N/A}
	\item Did you include complete proofs of all theoretical results?
    \answerNA{N/A}
\end{enumerate}

\item Additionally, if you ran machine learning experiments...
\begin{enumerate}
  \item Did you include the code, data, and instructions needed to reproduce the main experimental results (either in the supplemental material or as a URL)?
    \answerYes{Yes, the code and the anonymous user survey and interaction data are available at \url{https://github.com/IIT-ML/icwsm-2024-filter-bubbles}.}
  \item Did you specify all the training details (e.g., data splits, hyperparameters, how they were chosen)?
    \answerYes{Yes, see ``Our Approach'' section.}
     \item Did you report error bars (e.g., with respect to the random seed after running experiments multiple times)?
    \answerYes{Yes, please see Figures \ref{fig.measures} and \ref{fig.post_Qs}.}
	\item Did you include the total amount of compute and the type of resources used (e.g., type of GPUs, internal cluster, or cloud provider)?
    \answerNA{N/A}
     \item Do you justify how the proposed evaluation is sufficient and appropriate to the claims made? 
    \answerYes{Yes, please see ``Measures'' subsection in the Approach section (Section \ref{sec.details})}
     \item Do you discuss what is ``the cost`` of misclassification and fault (in)tolerance?
    \answerNA{N/A}
  
\end{enumerate}

\item Additionally, if you are using existing assets (e.g., code, data, models) or curating/releasing new assets...
\begin{enumerate}
  \item If your work uses existing assets, did you cite the creators?
    \answerYes{Yes, see the ``Datasets'' subsection of Section \ref{sec.details}.}
  \item Did you mention the license of the assets?
    \answerNA{NA}
  \item Did you include any new assets in the supplemental material or as a URL?
    \answerYes{Yes, the code and the anonymous user survey and interaction data are available at \url{https://github.com/IIT-ML/icwsm-2024-filter-bubbles}.}
  \item Did you discuss whether and how consent was obtained from people whose data you're using/curating?
    \answerYes{Yes, see ``Technical Appendix'' section. We also restate here that the study was granted IRB approval by the authors' institutions.}
  \item Did you discuss whether the data you are using/curating contains personally identifiable information or offensive content?
    \answerYes{Yes, please see the Appendix.}
\item If you are curating or releasing new datasets, did you discuss how you intend to make your datasets FAIR)?
\answerNA{N/A}
\item If you are curating or releasing new datasets, did you create a Datasheet for the Dataset)? 
\answerNA{N/A}
\end{enumerate}

\item Additionally, if you used crowdsourcing or conducted research with human subjects...
\begin{enumerate}
  \item Did you include the full text of instructions given to participants and screenshots?
    \answerYes{Yes, see ``Technical Appendix'' section.}
  \item Did you describe any potential participant risks, with mentions of Institutional Review Board (IRB) approvals?
    \answerYes{Yes.}
  \item Did you include the estimated hourly wage paid to participants and the total amount spent on participant compensation?
    \answerYes{Yes, please see the ``Compensation'' subsection in the appendix.}
   \item Did you discuss how data is stored, shared, and deidentified?
   \answerYes{Yes, please see Appendix.}
\end{enumerate}

\end{enumerate}

\appendix


\section*{Technical Appendix}
\label{sec:appendix}

\subsection{User recruitment} 
The study conducted was granted IRB approval by the authors' institutions. The study was conducted through Amazon Mechanical Turk and no personally identifiable information was collected at any time. The initial screening survey contained six demographic questions: gender, age, race, self-identified political stance, education level, and annual income. Additionally, we ask these individuals the following three questions:
\begin{enumerate}
    \item \textit{Where do you get most of your information about current news events?} Response options include \textit{printed, online, TV,} among others. 
    \item \textit{How often do you read or watch news about U.S. politics, policies, or the economy?} The response is a five-point Likert scale ranging from \textit{Never} to \textit{Always}.
    \item \textit{How often do you use fact-checking websites  (e.g., PolitiFact, Snopes, FactCheck, etc.)?} The response option is a five-point Likert scale ranging from \textit{Never} to \textit{Always}.
\end{enumerate}

The survey ends with a political literacy qualification section containing three questions meant to assess a basic knowledge of U.S.~politics. Those who answered at least two of the three following questions correctly are invited to participate in the full-scale recommendation study:
\begin{enumerate}
    \item \textit{Which of the following is the most conservative news source?} Response options are \textit{MSNBC, New York Times, Fox News, The Guardian}.
    \item \textit{Among the following, who is the most liberal politician?} Response options are \textit{Ted Cruz, Bernie Sanders, Donald Trump, Lindsey Graham.}
    \item \textit{Which state among the following recently enacted a restrictive abortion law?} Response options are \textit{Texas, Massachusetts, New York, California}. (The study was conducted immediately after Texas drafted its Texas Heartbeat Act in September 2021.) 
\end{enumerate}

\subsection{User interface} 
\begin{figure}[!ht]
    \centering
    \includegraphics[width=0.95\linewidth]{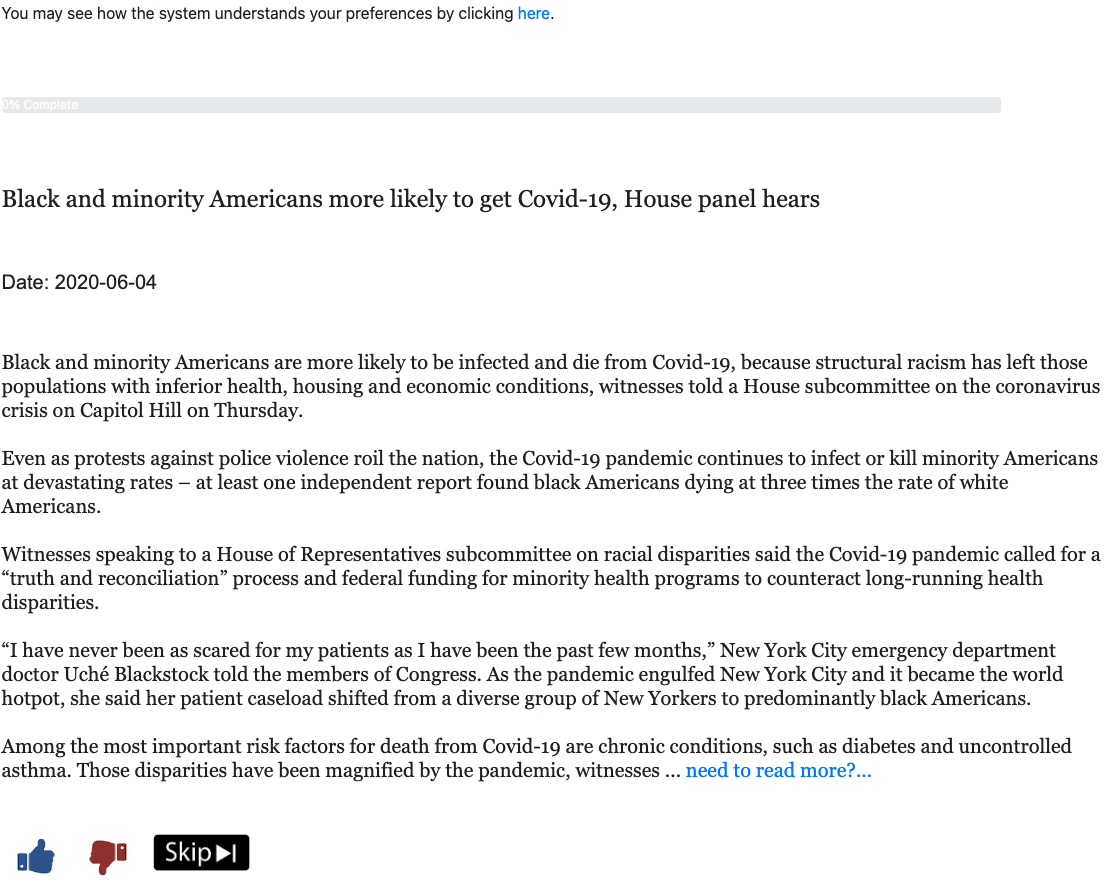}
    \caption{An example recommendation page for a user in the treatment group. The sentence at the top ``You may see how the system understands your preferences by clicking here,'' was available only to users in the treatment group, connecting them to the transparency and interaction tool.}
    \label{fig.recommendation.page}
\end{figure}

An example recommendation page is shown in Figure \ref{fig.recommendation.page}. Users have access to the title, date, and content of the article. At the bottom of the page are the up-vote, down-vote, and skip buttons. The example in Figure \ref{fig.recommendation.page} is for a user in the treatment group; those in the control group see a similar page with the sentence and hyperlink at the top (beginning with ``You may see how...'') removed.

\begin{figure}[t]
  \centering
    \includegraphics[width=0.48\textwidth]{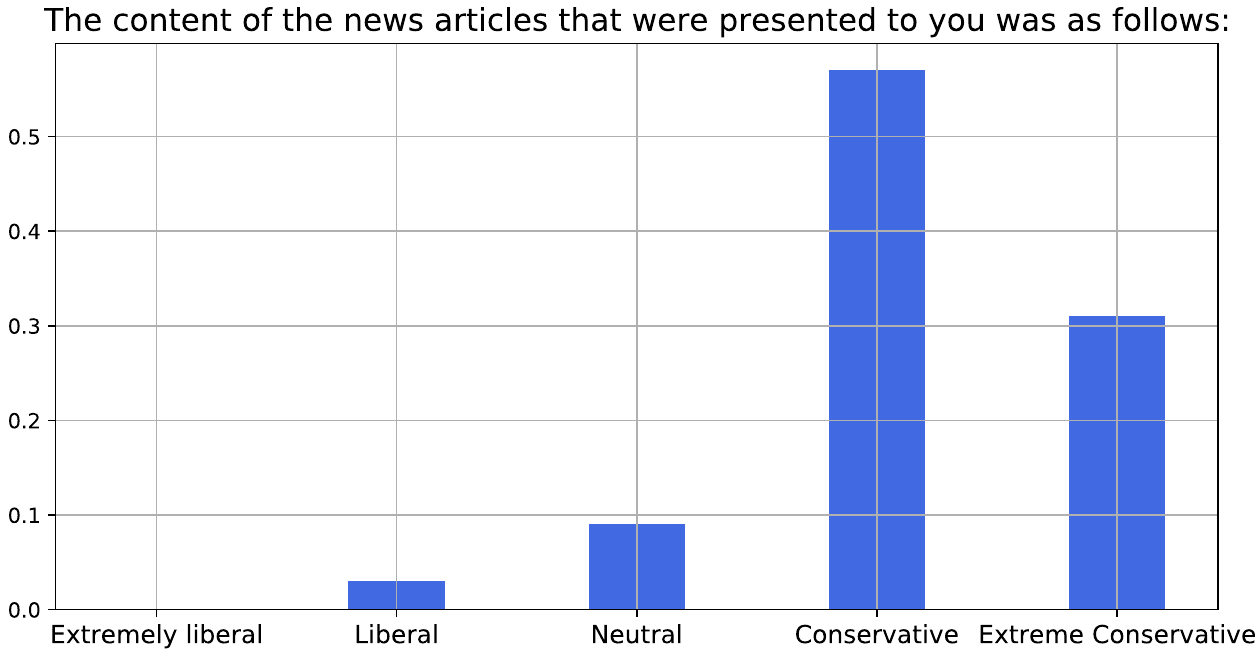}
  \caption{A sample transparency figure provided to users in the control group between answering \emph{Qb} and \emph{Qc.}} \label{fig.transparency}
\end{figure}

Figure~\ref{fig.transparency} shows a sample transparency figure shown to users in the control group.

\subsection{Instructions to users}

Here are the instructions users were provided at each stage.

\paragraph{Before pre-Questionnaire} {\em Thank you for your participation. Please take 3-5 minutes to answer the
following questions about American policy issues. Click the ``Begin'' button to start. }

\paragraph{Selecting interests} {\em Regardless of your previous responses, how interested would you be in reading news articles about the following topics?}

\paragraph{Before the first recommendation is shown} Both control and treatment users: {\em Thank you for your responses thus far. You will now assess a number of newspaper articles according to your personal preferences.
Your options after reading each article are to {\em give a thumbs-up} (you enjoyed reading the article), {\em give a thumbs-down} (you did not enjoy reading the article), or {\em skip} (you had no strong feelings about the article).} The users in the treatment group was shown the following additional two sentences {\em 
You will periodically be exposed to information about how the system understands your preferences about the news, and you will have opportunities to view and modify this information. Please click the button below to begin.}

\paragraph{On the interaction page (treatment group only)} {\em Below is a description of how the system understands your news-related preferences. 
You have two options.: 
1) You may move the ``Political Stance'' slider to receive more articles in your preferred stance. 
2) You may move the ``Interest'' slider to adjust the number of articles on a topic. 
When you have completed making changes, if any, click the SUBMIT button to read more news articles
After all changes have been made, click SUBMIT to continue reading newspaper articles.
You can revert to your original preferences based on your survey responses by clicking REVERT button.}

\paragraph{After all recommendations and before post-questionnaire} {\em Thank you for your participation.
Now you are going to answer the last several post-questionaires to get your unique token.
If you have any questions of this survey, please contact author-info-removed-for-anonymous-review.}

\paragraph{Post-questionnaire} {\em To what extent do you agree or disagree with the following statements.}

\paragraph{At the end of the study} {\em Thank you for your participation. Here is your hash string that you need to copy and paste in Amazon Mturk website. If you have any questions of this survey, please contact author-info-removed-for-anonymous-review.}

\subsection{Attention check} We implemented an attention-checking mechanism to ensure that users are not randomly clicking up/down-vote buttons. Ten articles focusing on science-related news, a politically neutral topic, had embedded in the article content instructions for the user to respond in a specific way regardless of their views, i.e. to up-vote, down-vote, or skip the article. Users who failed to click on the correct button on five articles or more were excluded from the study. Of the 146 users who participated, 44 failed this attention check and were removed from the study.

\subsection{Compensation} Participants were paid \$1 for completing the brief pre-screening survey and \$15 for completing the full user study. Based on the time spent in the study, we estimate this wage to be \$15/hour.

\end{document}